\documentclass[aps,prl,preprint,groupedaddress,letterpaper]{revtex4}
\usepackage{graphicx}
\usepackage{verbatim}
\newcommand{\bx}{\mathbf x}
\newcommand{\bv}{\mathbf v}

\newcommand{\bB}{\mathbf B}
\newcommand{\bJ}{\mathbf J}
\newcommand{\D}{\mathcal D}
\newcommand{\wpe}{\omega_{pe}}

\newcommand{\dadb}[2]{\frac{\partial #1}{\partial #2}}
\begin{document}
\title{Kinetic dissipation and anisotropic heating
in a turbulent collisionless plasma}
\author{T.~N.~Parashar}
\affiliation{Department of Physics \& Astronomy, 217 Sharp Lab,
University of Delaware, Newark, Delaware 19716, USA}

\author{M.~A.~Shay}
\affiliation{Department of Physics \& Astronomy, 217 Sharp Lab,
University of Delaware, Newark, Delaware 19716, USA}

\author{P.~A.~Cassak}
\affiliation{Department of Physics \& Astronomy, 217 Sharp Lab,
University of Delaware, Newark, Delaware 19716, USA}

\author{W.~H.~Matthaeus}
\affiliation{Department of Physics \& Astronomy, 217 Sharp Lab,
University of Delaware, Newark, Delaware 19716, USA}

\date{\today}
\begin{abstract}

  The kinetic evolution of the Orszag-Tang vortex is studied using
  collisionless hybrid simulations. In magnetohydrodynamics this
  configuration leads rapidly to broadband turbulence.  At small
  scales, differences from magnetohydrodynamics arise, as energy
  dissipates into heat almost exclusively through the magnetic
  field. A key result is that protons are heated preferentially in the
  plane perpendicular to the mean magnetic field, creating a proton
  temperature anisotropy as is observed in the corona and solar wind.
\end{abstract}
\maketitle

The dissipation of turbulent energy in plasmas plays a critical role
in understanding coronal heating and the acceleration of the solar
wind \cite{Cranmer02}, turbulence in the interplanetary medium
\cite{SmithEA06}, energy storage and release in the magnetosphere
\cite{SundkvistEA07}, and in a variety of other plasma and
astrophysical contexts \cite{Spangler03}. By ``dissipation,'' we mean
the conversion of fluid scale energy irreversibly into kinetic degrees
of freedom.
A key observational clue about the nature of this dissipation is the
substantial heating of protons, often preferentially in the plane
perpendicular to the mean magnetic field\cite{Cranmer02,TuMarsch95},
both in the solar wind (plasma $\beta=($thermal speed/Alfv\'en
speed)$^2 \sim 1$) and in the corona ($\beta <<1$).  This heating has
a variety of potential sources, including shocks and wave-particle
interactions involving non thermal distributions such as pickup ions,
but the ubiquity of broadband Kolmogoroff-like fluctuations suggests
that kinetic absorption of fluid energy at or beyond the high
wavenumber end of the inertial range plays an important role.

A fundamental demonstration of turbulent anisotropic heating 
is needed as a first step towards a basic physics
understanding of the dissipation processes
that heat the solar corona and solar wind. 
Magnetohydrodynamics (MHD) is a very useful 
plasma model which generally employs small but nonzero viscosity 
and/or resistivity, but this is not easily justified for 
collisionless systems where the mean free path is comparable to 
the system size (as in most of the corona and solar wind).  
More sophisticated attempts to numerically model
plasma dissipation, e.g., by employing hyperresistivity,
hyperviscosity or indirectly by including the Hall or finite Larmor
radius effects (see \cite{goldstein-chapter}), include only selected
approximations to kinetic effects, and therefore do not include the
wider range of mechanisms available to the kinetic plasma.

In this paper, we report a demonstration of turbulent anisotropic
proton heating in the Orszag-Tang vortex\cite{otv-ot} using a hybrid
simulation, which includes all proton kinetic effects.  The hybrid
simulation results are very similar at large length scales to MHD
simulations of the same system, but show significant differences at
small scales where kinetic effects are important.  Analysis of the
hybrid results show that energy is dissipated into proton heating
almost exclusively through the magnetic field and not through the
proton bulk velocity.  The proton heating occurs preferentially in the
plane perpendicular to the mean magnetic field. These simulations, to
our knowledge, are the first self consistent demonstration of
turbulent anisotropic proton heating.  Finally, effective transport
coefficients from the hybrid simulations are calculated, showing that
the approximation of constant resistivity $\eta$ is potentially
reasonable (although it cannot reproduce the proton temperature
anisotropy), but a constant viscosity $\nu$ is untenable. 

The Orszag-Tang vortex\cite{otv-ot} is a well studied MHD initial
configuration given by
\begin{equation}
{\mathbf B} \!  =  \! -\sin y~\hat{\mathbf x} \! + \!
    \sin 2x~\hat{\mathbf y} + B_g\, \hat{\bf z} ; \,
   \;\;{\mathbf v} \!  = \!  -\sin y~\hat{\mathbf x} \! + \! \sin x ~\hat{\mathbf y
}
\end{equation}
with ${\bf B}$ the magnetic field ($B_g$ a uniform guide field) and
${\bf v}$ the proton bulk velocity in normalized units described
later. This configuration leads immediately to strong nonlinear
couplings, producing cascade-like activity that might reasonably
approximate the highest wavenumber decade of the inertial range. These
couplings, which are dominantly local in wavenumber, in turn drive the
dissipation range.  The physics of the Orszag-Tang vortex has been
previously studied using incompressible \cite{otv-ot} and compressible
\cite{dahlburg} MHD simulations. Its robust production of nonlinear
activity is a motivation for its frequent use in validating numerical
schemes (see e.g. \cite{rosenberg2007}).

Simulating kinetic dissipation is difficult and computationally
expensive due to the requirement of treating a wide range of length
scales.  By choosing a computational domain with approximately one
decade of scale in the MHD range and another in the kinetic range, we
can study the conversion of strongly driven MHD fluctuations into
kinetic motions.  An antecedent of the present study compared global
behavior of hybrid and Hall MHD simulations \cite{matthews}, but
included a mean in-plane magnetic field while not adequately resolving
the proton inertial length.

We use the hybrid code P3D~\cite{shay-jgr-2001} in 2.5D, which models
protons as individual particles and electrons as a fluid and evolves
\begin{eqnarray}
	\frac{d \bx_i}{d t} & = & \bv_i \, ; \, \, \,
	\frac{d \bv_i}{d t}  =  \frac{1}{\epsilon_H}\left({\bf E}' + {\bf v}_i \times {\bf
          B}\right)\\
	\dadb{\bB'}{t} &=& \nabla\times\left(\bv\times\bB\right) -
	\epsilon_{H} \nabla\times\left(\frac{\bJ}{n}\times\bB'\right) 
	\label{induction} \\
	\bB' & \! = \! &
        \left(1 \! - \!  \frac{m_e}{m_i}\epsilon_{H}^2\nabla^2\right)\bB, 
        \;\;{\bf E}' \! = \! {\bf B} \times \left({\bf v} \!  - \! \epsilon_H \frac{\bf J}{n}\right)
	\label{bprimeeq}
\end{eqnarray}
where 
$\bJ=\nabla\times\bB$ is the current density,
$\epsilon_{H}\equiv c/(L_0 \omega_{pi})$ is the normalized proton
inertial length, $m_e$ and $m_i$ are the electron and proton masses,
$\bx_i$ and $\bv_i$ are the positions and velocities of the individual
protons, and $\bv$ is the proton bulk flow speed. Length is normalized to
$L_0,$ velocities to $V_0 = B_0/(4\pi m n_0)^{1/2}$, time to $t_0 =
L_0/V_0,$ and temperature to $B_0^2/(4\pi n_0).$ The average density is
$n_0$, and $B_0$ is the root mean square in-plane magnetic
field. The magnetic field ${\bf B}$ is determined from ${\bf B}'$ using
the multigrid method.  The code assumes
quasi-neutrality.  The electron temperature is zero and is not
updated.

Hybrid simulations are ideally suited for exploring dissipation and
proton heating in collisionless plasmas because they include a
complete kinetic description of protons. Due to the finite temperature
of the protons, kinetic Alfv\'en waves are present in this set of
hybrid simulations (\cite{Rogers01,Howes06} and references therein),
as well as parallel proton bulk flows.

\begin{figure}
  \includegraphics[width=7cm]{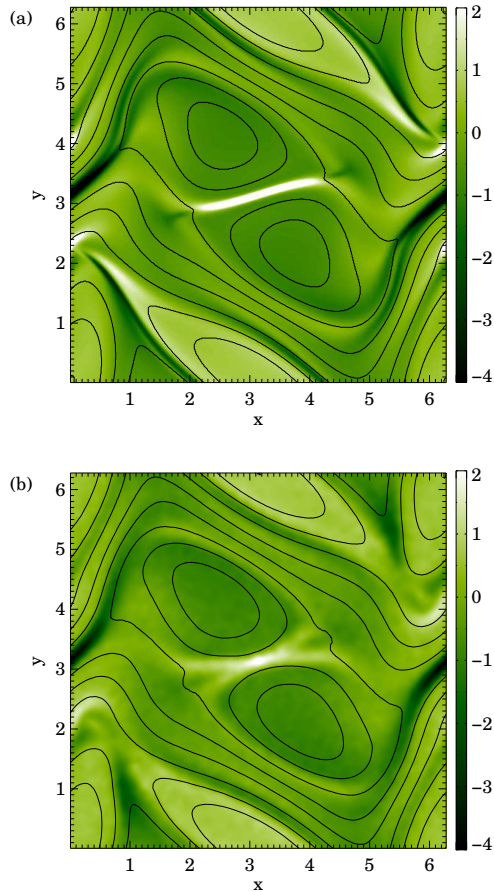}
  \caption{(Color online) Current density with magnetic flux contours
at $t$=1.96 in (a) fluid and (b) hybrid simulations.}
  \label{densities}
\end{figure}

The simulation domain is a square box of side length $2\,\pi \times
2\,\pi$ with $512 \times 512$ grid points. About 2.6 million protons
are loaded with an initial Maxwellian distribution having a uniform
temperature $=8$, and $\epsilon_{H}=2\pi/25.6$ and $m_e=0.04m_i$. No
artificial dissipation is present other than grid scale dissipation.
Choosing a guide field $B_g = 5$ (total $\beta = 2nT / B^{2} \approx
0.62$) reduces the system compressibility.  Incompressibility is
further promoted by adding perturbations to the background density
$n_0$ that enforce $\partial(\nabla\cdot\bv) / \partial t = 0$ at $t =
0$.  Simulations without the added perturbation show only small
differences.

The hybrid simulation results are compared to those of a compressible
2.5 D MHD version of the code F3D \cite{Shay04} with constant and
uniform resistivity $\eta = 0.0048$, zero viscosity $\nu$, and ratio
of specific heats $\gamma=5/3$. (We motivate values for $\eta$ and
$\nu$ later.)  In both cases, the magnetic islands initially centered
on the midplane ($y=\pi$) begin a clockwise rotation.  The initial
velocity profile shears the magnetic islands until $t \sim 2$ as the
islands approach and undergo a brief period of magnetic reconnection
from $t \sim 2 - 4$.  After $t \sim 4$, the system is dominated by
strong turbulence.  A comparison of out-of-plane current density $J_z$
and magnetic field lines at $t = 1.96$ is shown in
Fig.~\ref{densities}.  The hybrid and MHD results show strong
similarities at large scales, but significant differences at small
scales where kinetic effects become important.

\begin{figure}
  \includegraphics[width=8.5cm]{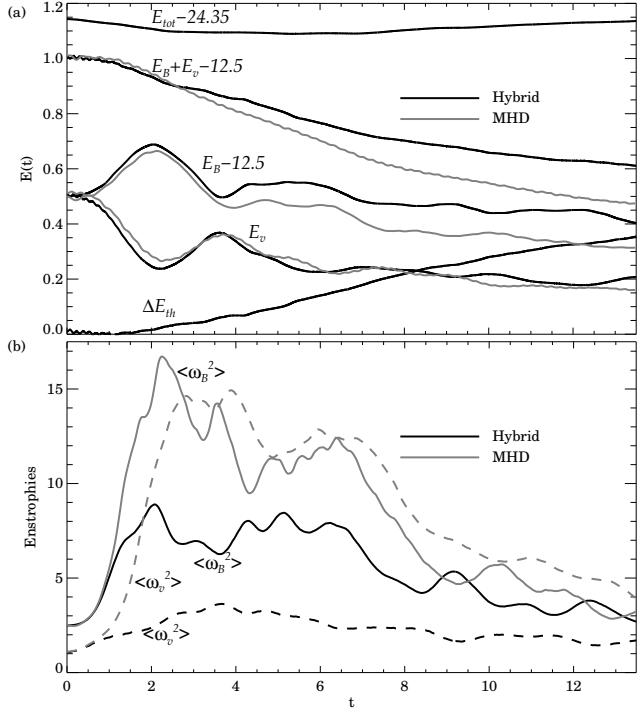}
  \caption{Hybrid and MHD comparison: (a) Magnetic energy 
$E_B$, fluid flow energy $E_v$, their sum, the change
in thermal energy $\Delta E_{th}$, and total energy $E_{tot}$
vs.~time. (b) Flow enstrophy $<\omega^2_v>$ and magnetic enstrophy
$<\omega_B^2 >$ vs.~ time.}
  \label{otv-pic}
\end{figure}

We can quantify the differences by comparing energy and dissipation
budgets, where flow energy $E_v = \langle \rho |{\bv}^2| / 2 \rangle$,
magnetic energy $E_B = \langle |{\bf B}|^2 / 2 \rangle$, thermal
energy $E_{th}$ (total proton kinetic energy minus flow energy), and
total energy $E_{tot}$.  $\langle \ldots \rangle$ denotes a
volume average.  Grid scale fluctuations in the hybrid data are
smoothed using a standard local, weighted iterative averaging.

Figure \ref{otv-pic}(a) shows $E_{v}, E_{B},$ their sum, $\Delta
E_{th}$ (where $\Delta$ means the change since $t = 0$) and $E_{tot}$,
as a function of time from the hybrid  and MHD simulations, with $E_B$
and $E_{tot}$
shifted down by a constant for convenience.  
Note that 
$E_{tot}$ changes very little over the course of the
hybrid run, demonstrating good numerical energy
conservation.  During the initial phase ($t < 2$), bulk
flow energy is converted strongly into magnetic energy
as field lines are stretched, but with little proton heating.  The
magnetic energy converts back to flow energy (with some heating) during
the reconnection event ($t \sim 2 - 4$).  Until $t \sim 4$, the
energetics of the hybrid and MHD results are very similar.  However,
in the turbulent phase ($t > 4$), the hybrid and MHD codes show
significant differences, and more dissipation occurs in the MHD case.
Notably, the proton thermal energy increases
monotonically during the turbulent phase in the hybrid simulation,
even without explicit dissipation. 

\begin{figure}
  \includegraphics[width=8.5cm]{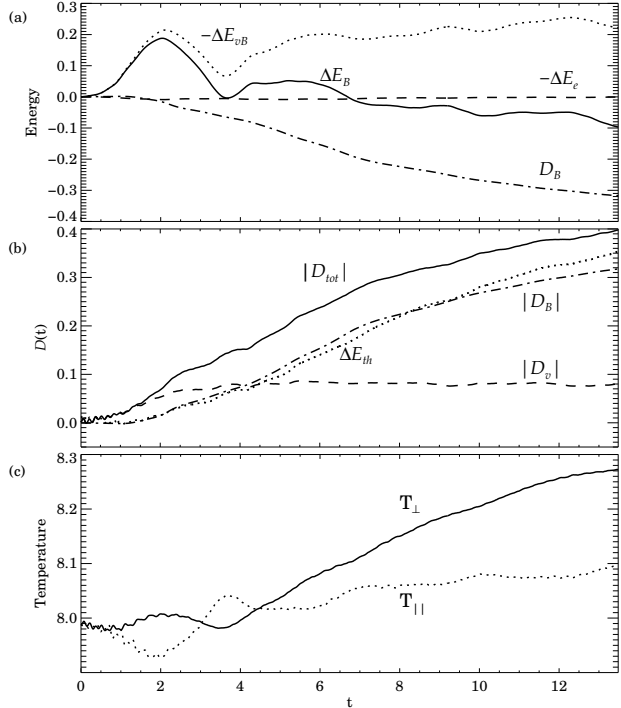}%
  \caption{(a) $\Delta E_{B}$: change of $E_B$ in the hybrid run; 
    $\Delta E_{vB}$: exchange
    between $E_v$ and $E_B$; $\Delta E_{e}$: 
    electron kinetic energy; $\D_{B}$: sum of these,  
   total $E_B$ dissipated. (b)
    $\D_v$ and $\D_B$ are cumulative dissipation through
    bulk flow and magnetic channels, $\D_{tot}$ their sum,
    $\Delta E_{th}$ change in thermal energy.  (c) Parallel
    and perpendicular proton temperatures vs.~time.}
  \label{d-plot}
\end{figure}

In MHD, mean square gradients of $\bf v$ and $\bf B$ are proportional
to the energy dissipation rate.  Although the hybrid code lacks
explicit dissipation, it is instructive to compare in
Fig.~\ref{otv-pic}(b) the out-of-plane enstrophy (mean square
vorticity) $\langle \omega_v^2 \rangle = \langle |\hat{\bf
  z}\cdot(\nabla \times {\bf v})|^2 \rangle$ and out-of-plane magnetic
enstrophy (mean square current density) $\langle \omega_B^2 \rangle =
\langle |\hat{\bf z}\cdot(\nabla \times {\bf B})|^2 \rangle = \langle
J_z^2 \rangle$ in the hybrid and MHD simulations. At early time ($t <
4$), the hybrid and MHD enstrophies peak at about same time, but their
magnitudes are very different, indicating that the length scales in
the hybrid case are larger, probably due to finite Larmor radius
effects. During the turbulent phase ($t > 4$), the enstrophies
continue to be different but the magnetic enstrophies are surprisingly
similar. This suggests that the kinetic dissipation may have some
resemblance to a classical resistivity, which we revisit later.  Note
that the value of enstrophy in the hybrid case is necessarily
sensitive to the averaging that defines the fluid scales.



To quantify the dissipation, consider the flow of
magnetic energy in the system.  Dotting the induction equation
[Eq.~(\ref{induction})] with ${\bf B}$, averaging over space,
and integrating over time gives
\begin{equation}
\Delta E_B(t) = -\int_0^t \langle
\bv\cdot(\bJ\times\bB) \rangle dt^{\prime} - \frac{d_e^2}{2}\langle
\Delta J^2(t) \rangle - \D_B(t),	\label{db-eqn}
\end{equation}
where $\Delta$ refers to the change since $t = 0$.  The
$\bv\cdot(\bJ\times\bB)$ term, called $\Delta E_{vB}$, is the
exchange in energy between bulk flow and
the magnetic field and the $d_{e}^{2}$ term is essentially the
electron kinetic energy.  We define $\D_B$ as the cumulative
energy dissipated from the magnetic channel.
Included in this term are non-MHD dissipative processes and grid
scale dissipation.
The first three terms are calculable from the simulations, 
so we may compute $\D_B$.  

Fig.~\ref{d-plot}(a) shows that for $t < 4$, $E_B$ 
increases due to input from $E_v$, 
then decreases as it is converted back.  
In the turbulent phase ($t > 4$), there
continues to be conversion into magnetic energy.  However, 
$E_B$ decreases during this period, 
and 
both energy converted from the flow, and
an approximately equal amount previously 
stored in the field, is absorbed in 
the magnetic dissipation $\D_B$.

A similar analysis can be done for energy flow in the bulk flow
channel.  Dotting the MHD momentum equation with ${\bf v}$ and
integrating over time and space gives
\begin{equation}
\Delta E_v(t) = \int_0^t \langle \bv\cdot(\bJ\times\bB) 
\rangle dt^{\prime} - \D_v(t),	\label{dv-eqn}
\end{equation}
where $\D_v$ is the cumulative energy converted into heat from the
flow channel through non-fluid effects and compression.
Fig.~\ref{d-plot}(b) shows $D_v$ (dashed line), $\D_B$ (dot-dashed),
$\D_{tot} = \D_B + \D_v$ (solid), and $\Delta E_{th}$ (dotted).
In the turbulent phase ($t > 4$), there is essentially no energy
dissipated through the flow channel ($\D_v$ is constant).  The
energy dissipated through the magnetic field $\D_B$ traces very closely the
increase in thermal energy of the protons $\Delta E_{th}$, 
so the main source of dissipation in
this system is through magnetic interactions. The small departure
between the two is accounted for by numerical heating, seen in the
change in total energy $E_{tot}$ [see Fig.~\ref{otv-pic}(a)];
this is only about 10 \% of the total dissipated magnetic energy

It is reasonable that little dissipation occurs through the flow
channel. In the MHD regime (low wavenumber $k$), the dynamics of the
hybrid simulation are at most weakly compressible, so the majority of
energy is in oblique Alfv\'en waves that are weakly damped
\cite{Gary99}.  As energy cascades to smaller scales, the proton
gyroradius is reached before the dissipation scale.  Below the proton
gyroradius, the ions decouple from the magnetic field and only weakly
participate with the non-MHD waves in this region.  Consequently, the
Alfv\'en ratio $E_v/E_B$ goes to zero in the kinetic regime as
evidenced by the structure of kinetic Alfv\'en and whistler waves.

\begin{figure}[t]
  \includegraphics[width=8.5cm]{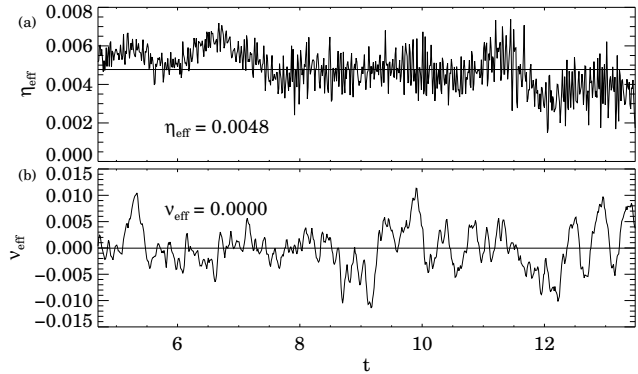}%
  \caption{\label{visc} Effective (a) resistivity $\eta$ and (b)
  viscosity $\nu$ vs.~time.}
\end{figure}

A central result of this study is that the dissipated magnetic energy
preferentially heats the protons perpendicular to the mean magnetic
field, as shown in Fig.~\ref{d-plot}(c). The perpendicular and
parallel temperatures $T_{\perp}$ and $T_{||}$ are calculated relative
to the guide field. In the turbulent phase ($t > 4$), $T_{\perp}$
increases monotonically, while $T_{||}$ remains relatively steady. The
relative anisotropy is small because the available magnetic free
energy in the system (from $B_x$ and $B_y$) is small compared to the
proton temperature, i.e. $\beta_\perp \gg 1.$ If the heating mechanism
in this study generalizes to $\beta_\perp \sim 1,$ large anisotropies
will develop. The perpendicular heating occurs without any obvious
connection to classical cyclotron resonances, as the latter generally
are construed \cite{IsenbergEA01} to involve gyroresonance with waves
propagating parallel to a background field (along the invariant
direction in this study). Resonance can also involve wave frequencies
near the cyclotron frequency; however, dominance in this simulation of
incompressible modes and relatively low frequency kinetic Alfv\'en
waves make the connection to standard cyclotron resonance
\cite{IsenbergEA01} uncertain.

By assuming the classical functional forms for the dissipation rates
$\eta \langle J_z^2 \rangle$ and $\nu \langle \omega_v^2 \rangle$, we
can compute effective transport coefficients $\eta_{{\rm eff}}$ and
$\nu_{{\rm eff}}$ from the hybrid simulations.  Figure \ref{visc}
shows $\eta_{{\rm eff}} = (\partial \D_B / \partial t) /\langle J_z^2
\rangle $ and $\nu_{{\rm eff}} = (\partial \D_v / \partial t) /\langle
\omega_v^2 \rangle$ versus time. Surprisingly, the spatially averaged
$\eta_{\rm eff}$ is fairly constant in time, as is often assumed in
MHD models.  The mean value is $\eta_{\rm eff} = 0.0048$,
corresponding to a magnetic Reynolds number of $S_{{\rm eff}} = 4 \pi
V_{0} L / \eta_{{\rm eff}} c^{2} \approx 1308$, which is the value
used in the MHD simulation. In classical turbulence theory, the length
scale at which dissipation occurs $\lambda_{d}$ is related to the
Reynolds number and energy containing scale $L$ through $S_{{\rm eff}}
\sim (L / \lambda_{d})^{4/3}$ \cite{Batchelor53}. For the hybrid
simulation, $\lambda_{d} \sim 0.029$, which is of the order of the
electron skin depth, $c / \wpe \sim 0.049$. The spatially averaged
$\nu_{\rm eff}$, on the other hand, shows oscillations much larger than
the mean, calling into question the assumption of a non-zero viscosity
assumed in many MHD models.

The effective resistivity being fairly constant does not imply that
the dissipation is of the form $\eta J^{2}$. The MHD simulations
performed with this $\eta_{\rm eff}$ show more dissipation of $B$ than
the hybrid simulations, and cannot reproduce the preferential heating
of $T_\perp.$ Future studies will investigate the spatial dependence
of $\eta_{\rm eff}$, and its dependence on electron/proton inertial
lengths, as well as system size. We will also examine the effect of
reducing the guide field and including  electron pressure in Ohm's
law. The physical mechanism which converts magnetic energy into proton
heat remains an open question and is under investigation. Potential
explanations are wave-particle interactions such as Landau damping.

This work is supported by NSF ATM-0539995, NASA NNG05GM98G and
NNX-08AI47G (Heliophysics Theory program), and NERSC. The authors
thank S. Servidio, S.P. Gary, and B. N. Rogers for enlightening
conversations.


\newcommand{\boldVol}[1] {\textbf{#1}} 
  \newcommand{\SortNoop}[1] {} 
  \newcommand{\au} {{A}{U}\ } 
  \newcommand{\AU} {{A}{U}\ } 
  \newcommand{\MHD} {{M}{H}{D}\ } 
  \newcommand{\mhd} {{M}{H}{D}\ } 
  \newcommand{\RMHD} {{R}{M}{H}{D}\ } 
  \newcommand{\rmhd} {{R}{M}{H}{D}\ } 
  \newcommand{\wkb} {{W}{K}{B}\ } 
  \newcommand{\alfven} {{A}lfv\'en\ } 
  \newcommand{\Alfven} {{A}lfv\'en\ } 
  \newcommand{\alfvenic} {{A}lfv\'enic\ } 
  \newcommand{\Alfvenic} {{A}lfv\'enic\ }

\end{document}